\documentclass[prc,aps,twocolumn,showpacs,preprintnumbers,nofootinbib,amsmath,amssymb]{revtex4}
\usepackage{psfig}
\usepackage{dcolumn}

\begin{document}
\title{Charmonium chemistry in A+A collisions at relativistic
energies}

\author{E.~L.~Bratkovskaya$^1$, A. P. Kostyuk$^{1,2}$,
W. Cassing$^3$ and H.~St\"ocker$^1$}
\affiliation{ \phantom{a}\\
 $^1$ Institut f\"{u}r Theoretische Physik,
      Universit\"{a}t Frankfurt, 60054 Frankfurt, Germany \\
 $^2$ Bogolyubov Institute for Theoretical Physics,
      03143 Kyiv, Ukraine \\
 $^3$ Institut f\"{u}r Theoretische Physik,
      Universit\"{a}t Giessen, 35392 Giessen, Germany}


\begin{abstract}
Charmonium production and suppression in heavy-ion collisions at
relativistic energies is investigated within different models, i.e.
the  comover absorption model, the threshold suppression model, the
statistical  coalescence model and the HSD transport approach.  In HSD
the charmonium dissociation cross sections with mesons are described by
a simple phase-space parametrization including an effective coupling
strength $|M_i|^2$ for the charmonium states $i=\chi_c, J/\psi,
\psi^\prime$.  This allows to include the backward channels for
charmonium reproduction by $D \bar{D}$ channels -- which are missed in
the comover absorption and threshold suppression model -- employing
detailed balance without introducing any new parameters.  It is found
that all approaches yield a reasonable description of $J/\psi$
suppression in S+U and Pb+Pb collisions at SPS energies. However, they
differ significantly in the $\psi^\prime/J/\psi$ ratio versus
centrality at SPS and especially at RHIC energies. These pronounced
differences can be exploited in future measurements at RHIC to
distinguish the hadronic rescattering scenarios from quark coalescence
close to the QGP phase boundary.
\end{abstract}



\maketitle


\section{Introduction}

The dynamics of ultra-relativistic nucleus-nucleus collisions at
Super-Proton-Synchrotron (SPS) and Relativistic-Heavy-Ion-Collider
(RHIC) energies are of fundamental interest with respect to the
properties of hadronic/partonic systems at high energy densities.
Especially the formation of a quark-gluon plasma (QGP) and its
transition to interacting hadronic matter has motivated a large
community for more than two decades \cite{Horst86,CaMo,Cass99,QM01}.
However,  the complexity of the dynamics has not been unraveled and the
evidence  for the formation of a QGP and/or the properties of the phase
transition is much debated \cite{QM04}. Apart from the light and
strange flavor ($u,\bar{u},d,\bar{d},s,\bar{s}$) quark physics and
their hadronic bound states in the vacuum ($p,n,\pi, K, \phi, \Lambda$
etc.), the interest in hadrons with charm ($c, \bar{c}$) has been
rising continuously since the heavy charm quark provides an additional
energy scale, which is large compared to $\Lambda_{QCD}$. The $c,
\bar{c}$ quark degrees of freedom are  of particular interest in
context with the phase transition to the  QGP since $c\bar{c}$ meson
states might no longer be formed due to color screening
\cite{Satz,Satznew}.

However, the suppression of $J/\psi$ and $\psi^\prime$ mesons in the
high density phase of nucleus-nucleus collisions at SPS energies
\cite{NA50aa,NA50bb,NA50b,NA50a,Ramello,NA38} might also be attributed
to a large extent to inelastic comover scattering (cf.
\cite{Capella1,Cass99,Vogt99,Gersch,Cass00,Kahana,Spieles,Gerland} and
Refs. therein) provided that the corresponding $J/\psi$-hadron cross
sections are of the order of a few mb
\cite{Gerland,Haglin,Konew,Ko,Sascha,Barnes1}. Theoretical estimates
here differ by more than an order of magnitude (cf.
\cite{Barnes1,Bernd,I3,korean} and Refs. therein), especially with
respect to $J/\psi$-meson scattering, such that the question of
charmonium suppression is not yet settled. On the other hand, at RHIC
energies further absorption mechanisms -- such as plasma screening and
gluon scattering -- might play a dominant role as suggested in Refs.
\cite{Kojpsi,Rappnew} and also lead to a substantial reduction of the
$J/\psi$ formation  in central Au+Au collisions.

Furthermore, it has been pointed out - within statistical models - that
at RHIC energies the charmonium formation from open charm and anticharm
mesons might become essential \cite{BMS,Goren,Kost_SPS1,Kost_SPS2} and
even exceed the yield from primary $NN$ collisions \cite{Rafelski}. One
of the prevailing questions thus is if open charm mesons and charmonia
will achieve thermal  equilibrium with the light mesons during the
nucleus-nucleus reaction. Furthermore, is the distribution of charm
(anti)quarks over open and hidden  charm mesons conform with the
statistical law at the same freeze-out parameters as anticipated in
Refs. \cite{BMS,Goren,Kost_SPS1,Kost_SPS2}.

In fact, a previous analysis within the Hadron-String-Dynamics (HSD)
transport model \cite{brat03} has demonstrated that the charmonium
production from open charm and anticharm mesons becomes essential in
central Au+Au collisions at RHIC. This is in accordance with
independent studies in Refs.  \cite{Rappnew,Ko}. On the other hand,
these backward channels have been found to be practically negligible
at SPS energies.
There is, however, an experimental claim \cite{NA50open} that open
charm might be enhanced by up to a factor of 3 in central
nucleus-nucleus collisions. In this case the hidden charm regeneration
processes might be essential already in (semi-)central collisions at
SPS energies \cite{Goren,Kost_SPS1,Kost_SPS2}. A possible reason of
the open charm enhancement is an increase of the effective production cross
sections of heavy quarks in the strongly interacting medium
\cite{Kost_open}. Also strong secondary meson-baryon channels
might be responsible for this enhancement as pointed out in Ref.
\cite{CaKo}. In short, there are presently more open questions than
solid answers.

Here we extend our previous studies \cite{brat03} with respect to
observable ratios of charmonium states, i.e. in particular the
$\psi^\prime/J/\psi$ ratio which is accessable by experiment.  We
compare the HSD results to the calculations within the standard
scenarios (including only suppression channels) as well as within the
statistical coalescence model (SCM) \cite{BMS}.

Our work is organized as follows: In Section II we remind the reader of
the standard models of charmonium suppression as well as of the
statistical coalescence model. We also briefly recall the 'input' of
the HSD transport approach with respect to charmonium and open charm
degrees of freedom.  In Section III the results of all models are
presented for S+U collisions at $\sqrt{s}=20$ GeV, for  Pb+Pb
collisions at $\sqrt{s}$ = 17.3 GeV and Au+Au collisions at $\sqrt{s}$
= 200 GeV. We present the yields of $J/\psi$ and $\psi^\prime$ as well
as their ratio as a function of centrality.  Section IV gives a summary
of our findings.

\section{Description of the models}

\subsection{The standard  models}

The standard approach to charmonium production in heavy-ion collisions
assumes that $c\bar{c}$ bound states  are created {\it only} at the
initial stage of the reaction in primary nucleon-nucleon collisions.
During the subsequent evolution of the system, the number of hidden
charm mesons is {\it suppressed} by i) the absorption of pre-resonance
charmonium states in nuclei (the normal nuclear suppression), ii) the
interactions of charmonia with secondary hadrons (comovers) and iii) a
possible dissociation of $c\bar{c}$ bound states in the deconfined
medium. The last mechanism was first expected in Ref.
\cite{Satz} and it was proposed, that charmonia might be used as a
probe for deconfinement in the state of matter created at the early stage of the
collision.

Two basic versions of the standard  scenarios have been
considered in the literature that both restrict to suppression mechanisms, only.
One of them, the comover model
\cite{Capella1}, assumes that the charmonium suppression increases
gradually with the density of the strongly-interacting medium created
in the collision. No abrupt changes of absorption properties of the
medium take place.
The  model of Ref. \cite{Blaizot1}
represents the opposite extreme: the suppression sets in abruptly as
soon as the energy density exceeds a threshold value, which is a free
parameter of this model. This version of the `suppression-only´
approach will be
referred to as `the threshold scenario'. The latter model is motivated by the
idea that the charmonium dissociation rate is drastically higher
in a quark-gluon-plasma (QGP)  than in a hadronic medium.

For a brief description of the `suppression-only´ approach let us
consider two nuclei $A$ and $B$ that collide at impact parameter $b$.
The number of produced hidden charm mesons is given  by \cite{Kharzeev}
\begin{equation}
N^{AB(b)}_i =  \sigma^{NN}_i
A B \int d^2 s T_A(|\vec{s}|) T_B(|\vec{s}-\vec{b}|)
S(\vec{b},\vec{s}), \label{Ni}
\end{equation}
where $\sigma^{NN}_i$ is the production cross section of the charmonium
species $i$ in nucleon-nucleon ($N+N$) collisions, $T_{A(B)}$ is the
nuclear thickness function related to the nucleon density in the
nucleus, and $S(\vec{b},\vec{s}) < 1$ is a factor responsible for the
charmonium suppression.

At the very initial stage charmonia experience  absorption,
$S=S^{abs}$, by interactions with nucleons of the colliding nuclei
(see, for instance, Refs. \cite{Kharzeev,Capella1}). Bound $c\bar{c}$
states are assumed to be absorbed in the so-called `pre-resonance
state' before the final hidden charm mesons are formed.  This
absorption cross section is therefore taken to be the same for all
charmonia.
The cross section $\sigma_{abs}=4.4$~mb \cite{Cortese} is taken from
the most recent SPS data analysis and is close to the theoretical
prediction of Ref. \cite{Gerland_PRL}.  We assume that the same cross
section $\sigma_{abs}$ prevails also at RHIC energies.

Those charmonia --- that survive normal nuclear suppression --- are
furthermore subjected to the comover
\cite{Capella1,Cass99,Vogt99,Gersch,Cass00},
\cite{Kahana,Spieles,Gerland}
or quark-gluon plasma (QGP) suppression  \cite{Blaizot1}.  We recall that
both suppression scenarios describe successfully the centrality
dependence of the $J/\psi$ yield in Pb+Pb collisions at the SPS.  In
the comover approach, an additional suppression factor appears:
$S=S^{abs} S^{co}$ \cite{Capella1}, which depends on the density of
comovers and on an {\it effective} cross section $\sigma_{co}$ for
charmonium dissociation by comovers.  The value
$\sigma_{co}^{J/\psi}=1.0$~mb is obtained from the fit of the NA50 data
on $J/\psi$ production in Pb+Pb at SPS (new  data \cite{Ramello} were
added).  It also agrees with the NA38 data for S+U collisions.
This value corresponds to an average over all comover species, relative
collision energies as well as all charmonium states contributing to the
$J/\psi$ yield through their decays.  From a fit of the $\psi'$ to
$J/\psi$ ratio in S+U collisions at SPS we get the value
$\sigma_{co}^{\psi'}=5$~mb for the effective cross section of $\psi'$
suppression. We will assume that the cross sections
$\sigma_{co}^{J/\psi,\psi'}$ are the same also at RHIC energies;
however, the charmonium suppression at RHIC becomes stronger due to the
higher comover density.

There are two reasons for an increased comover density at RHIC relative
to SPS: a)  The multiplicity of produced secondary hadrons per unit
rapidity interval at midrapidity increases by a factor of about 1.5
from $\sqrt{s}=17$ GeV to $\sqrt{s}=200$ GeV already in elementary
nucleon-nucleon collisions; b) the deviations from the wounded nucleon
model become stronger at higher energies, which increases the comover
density in central nucleus-nucleus collisions additionally.  The
centrality dependence of the number of light-flavored hadrons per unit
pseudorapidity interval in Au+Au collisions at RHIC can be parametrized
as \cite{KhN}
\begin{equation}\label{nch_AA}
\left. \frac{d N_{h}^{\mbox{\footnotesize AuAu}}}{d y} \right|_{y=0}  =
\left. \frac{d N_{h}^{pp}}{d y} \right|_{y=0}
\left[ (1 - x) N_p/2 + x N_{coll} \right]~,
\end{equation}
where $x=0.11$ for $\sqrt{s}=200$ GeV \cite{PhobosX}, $N_p(b)$ is the
number of participants and $N_{coll}(b)$ is the number of collisions.
Both are evaluated in the Glauber approach.

Calculating the centrality dependence of the charmonium suppression, it
is convenient to introduce an {\it reactant} density in
the plane transverse to the collision axis:
\begin{equation}\label{nstar}
n_p^{*}(\vec{b},\vec{s})=
\left[ (1 - x) n_p(\vec{b},\vec{s}) + 2 x n_c(\vec{b},\vec{s}) \right].
\end{equation}
Here $n_p(\vec{b},\vec{s})$ and $n_c(\vec{b},\vec{s})$ are,
respectively, the densities of nucleon participants
and collisions in the transverse plane:
\begin{equation}
N_p(b)=\int d^2s~ n_p(\vec{b},\vec{s}) \end{equation} and
\begin{equation}
N_{coll}(b)=\int d^2s~ n_{coll}(\vec{b},\vec{s}). \end{equation}
Note that the multiplicity of light-flavored hadrons (\ref{nch_AA})
is proportional to \begin{equation}
N_p^{*}(b)=\int d^2s~ n_p^{*}(\vec{b},\vec{s}). \end{equation}
Motivated by this fact, we assume that the comover {\it density}
in the transverse plane,
which is needed to calculate $S^{co}$, is proportional
to $n_p^{*}$ (\ref{nstar}).

In contrast to the co-mover version of the suppression models,
the threshold scenario \cite{Blaizot1} assumes that no charmonia
are destroyed by the medium until the energy density reaches a
threshold value.  The excited charmonia $\chi_c$, which contribute
about 40\% to the total $J/\psi$ yield, are suppressed at lower energy
densities compared to directly produced $J/\psi$'s.  We have updated the fit
\cite{Blaizot2} to the SPS data (new NA50 data \cite{Ramello} were added)
using the corrected value of the normal nuclear absorption cross section
$\sigma_{abs} = 4.4 \pm 0.5$ mb \cite{NA50_03}.
The $J/\psi$ to Drell-Yan ratio in nucleon-nucleon collisions
is aproximatelly the same as in Ref. \cite{Blaizot2}:
$\sigma^{NN}_{J/\psi}/\sigma^{NN}_{DY}\approx 53$.
Our results are \footnote{These numbers are different from those of
Ref. \cite{Kost_RHIC}, where another value of the $J/\psi$ to
Drell-Yan ratio in nucleon-nucleon collisions
$\sigma^{NN}_{J/\psi}/\sigma^{NN}_{DY}\approx 43$ was assumed.}
$n_{\chi}=2.0$ fm$^{-2}$ and
 $n_{J/\psi}=3.8$ fm$^{-2}$. Here
$n_{\chi}$ ($n_{J/\psi}$) is the participant density in the transverse plane
corresponding to the threshold energy density at which
$\chi_c$-charmonia ($J/\psi$'s) are fully suppressed.  (The change
of the $J/\psi$ yield due to the $\psi'$ suppression is neglected.)
The threshold for the $\psi'$ suppression $n_{\psi'}=1.7$ is obtained
from a fit of the $\psi'$ to $J/\psi$ ratio in S+U collisions at SPS.

Extrapolating to RHIC energies, one again has to take into account that
the number of produced hadrons per unit rapidity and, consequently, the
energy density of the produced medium grows with the collision energy
and centrality. Due to deviations from the wounded nucleon model
(\ref{nch_AA}) the charmonium suppression sets in when the {\it
effective} participant density  $n_p^{*}(\vec{b},\vec{s})$
(\ref{nstar}) -- rather than  $n_p(\vec{b},\vec{s})$ --
exceeds the threshold value.  The number of secondary hadrons per {\it
effective} participant pair at $\sqrt{s}=200$ is higher than that at
the SPS by a factor of about $1.5$.  The critical energy density at
RHIC is reached, therefore, at lower effective participant density:
$n_{\chi}^{*} = n_{\chi}/1.5 \approx 1.3$~fm$^{-2}$ and
$n_{J/\psi}^{*}= n_{J/\psi}/1.5. \approx 2.5$~fm$^{-2}$.
$n_{\psi'}^{*}= n_{\psi'}/1.5. \approx 1.1$~fm$^{-2}$.

\subsection{Statistical coalescence model}

In contrast to the suppression scenario described in Section IIA, the
statistical coalescence model \cite{BMS} assumes that hidden and open
charm hadrons are created at hadronization near the point of chemical
freeze-out which might be close to the phase boundary of the QGP.
However, contrary to the pure thermal model \cite{gago1} the
total amount of charm in the system is not assumed to be in chemical
equilibrium. Indeed the relaxation time for the number of $c$ and
$\bar{c}$ is expected to exceed the lifetime of the system.  Therefore,
the total charm content of the final hadron system is assumed to be
equal to the number of $c$ and $\bar{c}$ created at the initial stage
of A+A reactions by nucleon-nucleon collisions.  Only the distribution of
$c$ and $\bar{c}$ among different hadron states is controlled by
statistical laws  in terms of the hadron gas (HG) model parameters:
i.e. temperature $T$, baryonic chemical potential $\mu_b$ and volume
$V$. It appears that the number of hidden charm mesons produced by a
statistical coalescence mechanism depends weakly on the
thermodynamic hadronization parameters $T$ and $\mu_{B}$. The
charmonium yield is mainly defined by the average number of charmed
quark-antiquark pairs $\overline{N}_{c\bar{c}}$ and by the
hadronization volume parameter $V$. We recall that, if
$\overline{N}_{c\bar{c}}$ is not much larger than unity, a proper
account  for the exact charm conservation becomes essential as shown in
Ref. \cite{Goren}. This is crucial at  SPS energies, where
$\overline{N}_{c\bar{c}}$ is less than unity, and remains essential for
moderate centralities in Au+Au collisions at RHIC.

The SCM formula for the {\it total}\ \ ($4 \pi$)\ \ charmonium yield,
that takes into account exact conservation of the number of $c\bar{c}$
pairs, was obtained in Ref. \cite{Goren}. In the real experimental
situation, however, measurements are performed in a limited rapidity
window $\Delta y$.  In the most simple case, when the fraction of
charmonia in the relevant rapidity window does not depend on the
centrality, one can merely use the formula for the total yield
multiplied by some factor $\xi < 1$.  This approach was used in Refs.
\cite{Kost_SPS1,Kost_SPS2} for studying the SPS data, where the
multiplicity of light
hadrons, which determine the freeze-out volume of the system, are
approximately proportional to the number of nucleon participants $N_p$
at all rapidities.  At RHIC the situation is different: the {\it
total} ($4\pi$) multiplicity of light hadrons are approximately
proportional to the number of participants $N_{part}$, while {\it at
midrapidity} it grows faster with $N_{part}$ [see Eq.(\ref{nch_AA})].
The centrality dependence of charmonium production at different
rapidities should, in this case, be also different.  To compare the SCM
prediction to the PHENIX data \cite{PHENIX_AA}, which are related to
the $J/\psi$ yield at midrapidity $d N_{J/\psi}/dy$, one has to derive
a formula for the charmonium yield in a {\it finite} rapidity interval
$\Delta y$.

To this aim let $\xi_{\Delta y} < 1$ be the probability that a $c$ quark, produced
in a nucleus-nucleus collision, has rapidity $y$ within the interval
$\Delta y$.  The probability distribution of the number $k_c$ of $c$
quarks inside the interval $\Delta y$ for events with fixed {\it total}
($4 \pi$) number $N_{c\bar{c}}$ of $c\bar{c}$ pairs is given by the
binomial law:
\begin{equation}
f(k_{c}|N_{c\bar{c}}) =
\frac{N_{c\bar{c}}!}{k_{c}!~(N_{c\bar{c}} - k_{c})!}~\xi_{\Delta
y}^{k_{c}}~ (1-\xi_{\Delta y} )^{N_{c\bar{c}}-k_{c}}.
\end{equation}
The probability distribution of the number $k_{\bar{c}}$ of $\bar{c}$'s
inside the interval $\Delta y$ is assumed to be {\it independent} of
$k_c$ \footnote{This differs from Ref.\cite{theory1}, where an exact
equality, $k_{c}=k_{\bar{c}}$, within the chosen interval $\Delta y$ is
assumed. In fact, the net charm is exactly zero only in the total
system. In any finite rapidity interval, however, event-by-event
fluctuations with $k_{c} \not = k_{\bar{c}}$ are possible.}.  It
conforms to the same binomial law.  Event-by-event fluctuations of the
number of $c\bar{c}$ pairs $N_{c\bar{c}}$, created at the early stage
of $A+A$ reaction in independent nucleon-nucleon collisions are Poisson
distributed:
\begin{equation}
P(N_{c\bar{c}};\overline{N}_{c\bar{c}}) =
\exp \left(- \overline{N}_{c\bar{c}} \right)~
 \frac{\left( \overline{N}_{c\bar{c}} \right)^{N_{c\bar{c}}}}
 {N_{c\bar{c}}~!}.
\end{equation}
The probability of $c\bar{c}$ coalescence is proportional to the
product of their numbers and inversely proportional to the system
volume.  The proportionality coefficient depends on the thermal
densities of the open and hidden charm hadrons, and is the same as in
the case of the total charmonium yield \cite{Goren}.

The average multiplicity of the charmonium species $i$
at fixed values of $k_{c}$ and $k_{\bar{c}}$ is therefore
given by  \cite{Vill}
\begin{equation}\label{fix}
N_i^{\Delta y}{(k_{c} k_{\bar{c}})}  \approx k_{c}
k_{\bar{c}} \frac{n_{i}^{tot}}{(n_O/2)^2}~ \frac{1}{V_{\Delta y}}.
\end{equation}
In deriving Eq.(\ref{fix}) we used the fact that the
thermal number of hadrons with hidden charm  is much smaller than
that with open charm.
Folding Eq.(\ref{fix}) with the  binomial and Poisson distributions one
gets
\begin{equation}  \label{ch_i}
N^{\Delta y}_i  \approx \xi_{\Delta y}^2
\overline{N}_{c\bar{c}} \left(\overline{N}_{c\bar{c}} + 1\right)
\frac{n_{i}^{tot}}{(n_O/2)^2}~ \frac{1}{V_{\Delta y}}~,
\end{equation}
where $n_{O}$ is the thermal density of all open charm hadrons and
$n_{i}^{tot}$ is the total thermal density of the charmonium species
$i$ (including the decay contributions from the higher charmonium
states).  Both $n_{O}$ and $n_{i}^{tot}$ are calculated in the grand
canonical ensemble with the QGP hadronization parameters
$T,\mu_{B},V_{\Delta y}$ found from fitting the data of light-flavored
\footnote{At RHIC the strangeness as well as all other conserved
charges, excluding charm, can be safely considered in the grand
canonical ensemble.} hadron yields in the rapidity interval $\Delta y$.
The average number of $c\bar{c}$ pairs $\overline{N}_{c\bar{c}}$ is,
however, related to their {\it total} ($4 \pi$) yield.

The distinctive feature of the statistical coalescence model is
that the {\it ratio} of multiplicities of different charmonium
species is the same as in the equlibrium hadron gas. Therefore
the $\psi'$ to $J/\psi$ ratio
is practically independent on the centrality and only slightly
depends on the collision energy (due to the change of freeze-out
parameters).

In Au+Au collisions at $\sqrt{s}=200$~GeV the yield of
light-flavored hadrons at midrapidity  is fitted within the hadron
gas model with $T=177$~MeV and $\mu_{B}=29$~MeV \cite{BMSR}.
The centrality dependence of the volume is calculated from
\begin{equation}
V_{\Delta y=1} =
\frac{1}{n_{ch}(T,\mu_{B})}~
1.2~ \frac{d N_{ch}^{\mbox{\footnotesize AuAu}}}{d \eta}~,
\end{equation}
the coefficient 1.2 is needed to recalculate the
number of particles per unit {\it pseudorapidity} ($\eta$) interval
to that per unit {\it rapidity} ($y$) interval \cite{eta-y}.
Here $n_{ch}$ is the charged hadron density calculated in the HG model.

The
average number of the initially produced $c\bar{c}$
pairs in our calculations
is proportional to the number of
binary nucleon-nucleon collisions:
\begin{equation}\label{chmult}
\overline{N}_{c\bar{c}}= N_{coll}(b)
\sigma^{NN}_{c\bar{c}}/\sigma^{NN}_{inel}.
\end{equation}
Our statistical coalescence model calculations are done under the
assumption that the open charm multiplicity is enhanced
in nucleus-nucleus collisions at the SPS according to the experimental
claim in Ref. \cite{NA50open}; therefore the effective charm production
cross section $\sigma^{NN}_{c\bar{c}}$ in Eq.(\ref{chmult}) is assumed
to be larger by a factor of $\sim 3.5$ than in elementary
nucleon-nucleon collisions.  This 'enhancement' factor is expected to
become weaker at larger collision energies \cite{Kost_open}. Therefore,
we neglect it in our calculations for RHIC.

The charm production cross section, $\sigma^{NN}_{c\bar{c}}$, has been
measured at RHIC by the PHENIX Collaboration \cite{Averbeck}.  The
result is consistent with PYTHIA calculations:
$\sigma^{NN}_{c\bar{c}}\approx 650\ \mu$b.  This gives
$\bar{N}_{c\bar{c}} \approx 16-17$ for central Au+Au collisions in line
with the HSD calculations in Ref.  \cite{Cass01}.

The SCM is applicable only to large systems:  $N_{part} > 100$ in Pb+Pb
at the SPS \cite{BMS,Kost_SPS1,Kost_SPS2}.  Therefore, the PHENIX's
$p+p$ point and the most peripheral Au+Au point, corresponding to
$N_{part} \approx 30$, cannot be used in the SCM fit procedure. For
this reason we restrict ourselves only to a rough estimate of the SCM
prediction for the charmonium yield at midrapidity at the top RHIC
energy.

We fix the charm production cross section in nucleon-nucleon collisions
at its PYTHIA value, $\sigma^{NN}_{c\bar{c}} = 650\ \mu$b.  Since there
are no experimental data for the value of $\xi_{\Delta y =1}$,  one can
roughly estimate it assuming approximately the same rapidity
distribution for the open charm and $J/\psi$'s in $p+p$ collisions.
This leads to $\xi_{\Delta y =1} \approx 0.3$.  We note that the charm
rapidity distribution in Au+Au collisions might be broader than in
$p+p$ reactions due to rescattering of $c$ and $\bar{c}$ with nucleons.
This will not change our result essentially, however.  The estimate of
the total charm production cross section is based on the single
electron measurement at midrapidity.  Any extrapolation to the total
phase space has been done assuming that the charm rapidity distribution
does not change from $p+p$ to Au+Au. The charm production rate per
binary collision at midrapidity was found to be independent of the
centrality (at least within the present accuracy of the measurement).
This implies that the total charm production cross section should grow
with the centrality, if there is a broadening of the rapidity
distribution. Both effects, the decrease of $\xi_{\Delta y =1}$ and the
increase of $\sigma^{NN}_{c\bar{c}}$ nearly cancel each other in Eq.
(\ref{ch_i}) such that the prediction of SCM does not change
significantly.

\subsection{Open charm and charmonium dynamics in HSD}

In order to examine the dynamics of open charm and charmonium degrees
of freedom during the formation and expansion phase of the highly
excited system created in a relativistic nucleus-nucleus collision
within transport approaches, one has to know the number of initially
produced particles with $c$ or $\bar{c}$ quarks, i.e. $D, \bar{D}, D^*,
\bar{D}^*, D_s, \bar{D}_s, D_s^*, \bar{D}_s^*,$ $J/\psi(1S),
\psi^\prime(2S), \chi_c(1P)$.  In this work we follow the previous
studies in Refs.  \cite{Cass99,Cass00,Cass01,brat03} and fit  the total
charmonium production cross sections ($i = \chi_c,  J/\psi,
\psi^\prime$) from $NN$ collisions as a function of the invariant
energy $\sqrt{s}$ by the function
\begin{eqnarray}
\sigma_i^{NN}(s) = f_i \ a  \left(1 - \frac{m_i}{\sqrt{s}}\right)^\alpha
\left(\frac{\sqrt{s}}{m_i}\right)^\beta \theta(\sqrt{s}-\sqrt{s_{0i}}),
 \label{fitj}
\end{eqnarray}
where $m_i$ denotes the mass of charmonium $i$ while
$\sqrt{s_{0i}}=m_i+2 m_N$ is the threshold in vacuum.
The parameters in (\ref{fitj}) have been fixed to describe the  $J/\psi$
and $\psi^\prime$ data at lower energy ($\sqrt{s} \leq$ 30 GeV) as well
as the data point from the PHENIX Collaboration \cite{PHENIX_pp}
at $\sqrt{s}=200$ GeV, which gives $\sigma(pp\to J/\psi +X) =
3.99\pm 0.61(\rm{stat.})\pm 0.58(\rm{sys.}\pm 0.40 (\rm{abs}))$ $\mu$b
for the total $J/\psi$ cross section.
We use $a=0.2$ mb, $\alpha$ = 10, $\beta =0.775$.
The parameters $f_i$ are the fraction of charmonium states $i$.
For the present study we choose
$f_{\chi_c}=0.636, \ f_{J/\psi}=0.581,\ f_{\psi^\prime}=0.21$ in order
to reproduce the experimental ratio
$$\frac{B(\chi_{c1}\to J/\psi)\sigma_{\chi_{c1}}
 +B(\chi_{c2}\to J/\psi)\sigma_{\chi_{c2}}}
 {\sigma^{exp}_{J/\psi}}=0.344\pm 0.031$$
measured in $pp$ and $\pi N$ reactions \cite{E705_93,WA11_82} as well as the
averaged $pp$ and $pA$ ratio
$$(B_{\mu\mu}(\psi^\prime)\sigma_{\psi^\prime})
 / (B_{\mu\mu}(J/\psi)\sigma_{J/\psi})\simeq 0.0165$$
(cf. the compilation of experimental data in Ref. \cite{NA50_03}). Here
the experimentally measured $J/\psi$ cross section includes the direct
$J/\psi$ component $(\sigma_{J/\psi})$ as well as the decays of higher
charmonium states $\chi_{c}, \psi^\prime$, i.e.

\begin{eqnarray}
\sigma^{exp}_{J/\psi}=\sigma_{J/\psi}+B(\chi_{c}\to J/\psi)\sigma_{\chi_{c}}
+B(\psi^\prime\to J/\psi)\sigma_{\psi^\prime}. \
\label{xsexp}\end{eqnarray}
Note, we do not distinguish here the $\chi_{c1}(1P)$
and $\chi_{c2}(1P)$ states. Instead, we use only the $\chi_{c1}(1P)$ state
(which we denote as $\chi_c$), however, with an increased  branching ratio
for the decay to $J/\psi$ in order to include the contribution of
$\chi_{c2}(1P)$, i.e.  $B(\chi_{c}\to J/\psi) = 0.54$.
We adopt $B(\psi^\prime\to J/\psi)=0.557$ from \cite{PDG}.

For the total charmonium production cross sections from $\pi N$ reactions
we use the parametrization
(in line with Ref. \cite{Vogt99}):
\begin{eqnarray}
\sigma_i^{\pi N} (s) = f_i \ b \ \left(1 - \frac{m_i}{\sqrt{s}}\right)^\gamma
 \theta(\sqrt{s}-\sqrt{s_{0i}}),
\label{fitpin}\end{eqnarray}
with $\gamma=7.3$ and $b=1.24$~mb, which describes the
existing experimental data at low $\sqrt{s}$ reasonably well
(cf. Fig. 3 from \cite{Cass01}).  $\sqrt{s_{0i}}=m_i+m_N+m_\pi$
is the threshold in vacuum for $\pi N$ reactions.

Apart from the total cross sections, we also need the differential
distribution of the produced mesons in the transverse momentum $p_T$
and the rapidity $y$ (or Feynman $x_F$) from each individual collision.
We recall that $x_F = p_z/p_z^{max} \approx 2 p_z/\sqrt{s}$ with $p_z$
denoting the longitudinal momentum. For the differential distribution
in $x_F$ from $NN$ and $\pi N$ collisions we use the ansatz
from the E672/E706 Collaboration \cite{E672}:
\begin{equation}
\frac{dN}{dx_F dp_T} \sim (1 - |x_F|)^c \ \exp(-b_{p_T} p_T),
\label{fit2}
\end{equation}
where $b_{p_T}=2.08$ GeV$^{-1}$ and $c= a/(1+b/\sqrt{s})$. The
parameters $a, b$ are choosen as $a_{NN}=13.5$, $b_{NN}=24.9$ for
$NN$ collisions and $a_{\pi N}=4.11$, $b_{\pi N}=10.2$ for $\pi N$
collisions as in \cite{Cass01,brat03}.

The total and differential cross sections for open charm mesons
from $pp$ collisions, furthermore, are taken as in Refs.
\cite{Cass01,brat03}.  We thus refer to the results of Ref.
\cite{Cass01} which give $\sim$16 $D\bar{D}$ pairs in central
Au+Au collisions at $\sqrt{s}$ = 200 GeV, a factor of $\sim$160
relative to the expected primordial $J/\psi$ multiplicity.

Apart from primary hard $NN$ collisions the open charm mesons or
charmonia may also be generated by secondary meson-baryon ($mB$)
reactions. Here we include all secondary collisions of mesons with
baryons by assuming that the open charm cross section (from Section 2
of Ref. \cite{Cass01}) only depends on the invariant energy $\sqrt{s}$
and not on the explicit meson or baryon state.  Furthermore, we take
into account all interactions of 'formed' mesons -- after a formation
time of $\tau_F$ = 0.8 fm/c (in their rest frame) \cite{Geiss} -- with
baryons or diquarks, respectively. As pointed out in Ref. \cite{Cass01}
the production of open charm pairs in central Au+Au collisions by
$mB$ reactions at RHIC energies is expected to be on the 10\% level.

In order to study the effect of rescattering we tentatively adopt
the following dissociation cross sections of charmonia with
baryons independent on the energy (in line with Refs. \cite{Cass00,Cass01}):
\begin{eqnarray}
&& \sigma_{c\bar{c}B} = 6 \ {\rm mb}; \label{sigmacB} \\
&&\sigma_{J/\psi B} = 4 \ {\rm mb}; \ \sigma_{\chi_c B} = 5 \ {\rm mb};
\  \sigma_{\psi^\prime B} = 5 \ {\rm mb}.
\nonumber\end{eqnarray}
In (\ref{sigmacB}) the cross section $\sigma_{c\bar{c}B}$ stands for a
(color dipole)  pre-resonance ($c\bar{c})$ - baryon cross section,
since the $c\bar{c}$ pair produced initially cannot be identified with
a particular hadron due to the uncertainty relation in energy and time.
For the lifetime of the pre-resonance $c\bar{c}$ pair (in it's rest
frame) a value of $\tau_{c\bar{c}}$ = 0.3 fm/c is assumed following
Ref. \cite{Kharz}. This value corresponds to the mass difference of the
$\psi^\prime$ and $J/\psi$.

For $D, D^*, \bar{D}, \bar{D}^*$ - meson ($\pi, \eta, \rho, \omega$)
scattering we address to the calculations from Ref.  \cite{Konew,Ko}
which predict elastic cross sections in the range of 10--20 mb
depending on the size of the formfactor employed. As a guideline we use
a constant cross section of 10 mb for elastic scattering with mesons
and also baryons, although the latter might be even higher for very low
relative momenta.

As already pointed out in the Introduction the $J/\psi$ formation
cross sections by open charm mesons or the inverse comover
dissociation cross sections are not well known and the
significance of these channels is discussed controversely in the
literature \cite{Bernd,BMS,Rafelski,Redlich,I2,I3}. We here
follow the concept of Ref. \cite{brat03} and introduce a simple
2-body transition model with a single parameter $M_i^2$ for each
charmonium, that allows to implement the backward reactions
uniquely by employing detailed balance for each individual
channel. Since the meson-meson dissociation and backward reactions
typically occur with low relative momenta ('comovers') it is
legitimate to write the cross section for the process $1+2\to 3+4$ as
\begin{equation}
\label{model}
 \sigma_{1+2\to 3+4}(s) = 2^4 \frac{E_1 E_2 E_3 E_4}{s}
|\tilde M_i|^2 \left(\frac{m_3+m_4}{\sqrt{s}}\right)^6  \frac{p_f}{p_i},
\end{equation}
 where $E_k$ and $S_k$ denote the energy and spin of hadron $k$
$(k=1,2,3,4)$, respectively. The initial and final momenta for fixed
invariant energy  $\sqrt{s}$ are given by
\begin{eqnarray}
p_i^2 = \frac{(s-(m_1+m_2)^2)(s-(m_1-m_2)^2)}{4s}, \nonumber\\
p_f^2 = \frac{(s-(m_3+m_4)^2)(s-(m_3-m_4)^2)}{4s},
\label{moment}
\end{eqnarray}
where $m_k$ denotes the mass of hadron $k$. In (\ref{model}) $|\tilde
M_i|^2$ ($i=\chi_c, J/\psi, \psi^\prime$) stands for the effective
matrix element squared  which for the different 2-body channels is
taken of the form
\begin{eqnarray}
&&\hspace*{-3mm}|\tilde M_i|^2 =|M_i|^2  \ \ {\rm for} \
    \ (\pi,\rho)+(c\bar c)_i \to D+\bar{D} \label{mod}\\
&&\hspace*{-3mm}|\tilde M_i|^2 = 3 |M_i|^2  \ \ {\rm for} \
    \ (\pi,\rho)+(c\bar c)_i  \to D^*+\bar{D},\nonumber\\
&&\hspace*{-3mm} \hspace*{5cm} D+\bar{D}^*, \ D^* + \bar{D}^* \nonumber\\
&&\hspace*{-3mm}|\tilde M_i|^2 = \frac{1}{3} |M_i|^2 \ \ {\rm for}\
    \ (K,K^*)+(c\bar c)_i  \to D_s + \bar{D}, \nonumber\\
&&\hspace*{-3mm} \hspace*{6cm}  \bar{D}_s + D \nonumber \\
&&\hspace*{-3mm}|\tilde M_i|^2 =  |M_i|^2  \ \ {\rm for} \
    \ (K,K^*)+(c\bar c)_i  \to D_s + \bar{D}^*, \nonumber\\
&&\hspace*{2cm} \bar{D}_s + D^*,\ D^*_s + \bar{D},\  \bar{D}^*_s + D, \ \bar{D}^*_s + D^*
\nonumber
\end{eqnarray}
The relative factors of 3 in (\ref{mod}) are guided by the sum rule
studies in \cite{korean} which suggest that the cross section is
increased whenever a vector meson $D^*$ or $\bar{D}^*$ appears in the
final channel while another factor of 1/3 is introduced for each $s$ or
$\bar{s}$ quark involved. The factor $\left( {(m_3+m_4)}/{\sqrt{s}}
\right)^6 $ in (\ref{model}) accounts for the suppression of binary
channels with increasing $\sqrt{s}$ and has been fitted to the
experimental data for the reactions $\pi + N \rightarrow \rho+N,
\omega+N, \Phi+N, K^+ +\Lambda$ in Ref. \cite{CaKo}.

In Ref. \cite{brat03} we have used (for simplicity) the same matrix
elements for the dissociation of all charmonium states $i$ ($i=\chi_c,
J/\psi, \psi^\prime$) with mesons.  However, there is no fundamental
reason why these matrix elements should be identical. In the present
study we will explore the charmonium "chemistry" explicitly and
consider two different scenarios -- set 1:  the same matrix element for
all charmonium states $i$ as in Ref. \cite{brat03}, and set 2: the matrix
element squared for $\psi^\prime$ is enhanced by factor of 1.5
relative to $J/\psi$:
\begin{eqnarray}
{\rm set} 1: & |M_{J/\psi}|^2=|M_{\chi_c}|^2=|M_{\psi^\prime}|^2
            =|M_0|^2 \label{set12} \\
{\rm set} 2: & |M_{J/\psi}|^2=|M_{\chi_c}|^2 =|M_0|^2,\ \
               |M_{\psi^\prime}|^2 = 1.5 \ |M_0|^2.
\nonumber
\end{eqnarray}
We have fixed the parameter $|M_0|^2$ by comparison to the $J/\psi$
suppression data from the NA38 and NA50 Collaborations for S+U and Pb+Pb
collisions at 200 and 160 A$\cdot$GeV, respectively
\cite{NA38,NA50b,NA50a} (cf. Fig. \ref{Fig_JPSPS} in Section III).
We obtain the best fit for $|M_0|^2=0.17$~fm/GeV$^2$ (which is
slightly higher than in our previous study \cite{brat03} since the
fractions of charmonium states $f_i$ have been also modified here).

The advantage of the model introduced in (\ref{model}) is that
detailed balance for the binary reactions can be employed
strictly for each individual channel, i.e.
\begin{eqnarray}
\!\!\sigma_{3+4 \rightarrow 1+2}(s) =
 \sigma_{1+2 \rightarrow 3+4}(s)
\frac{(2S_1+1)(2S_2+1)}{(2S_3+1)(2S_4+1)} \ \frac{p_i^2}{p_f^2}, \
\label{balance}
\end{eqnarray}
 and the role of the backward reactions
($(c\bar c)_i$+meson formation by $D+\bar{D}$ flavor exchange) can be
explored without introducing any additional parameter once $|M_i|^2$
is fixed. The uncertainty in the cross sections (\ref{model}) is
of the same order of magnitude as that in Lagrangian approaches
using e.g. $SU(4)_{flavor}$ symmetry \cite{Konew,Ko} since the
formfactors at the vertices are essentially unknown \cite{korean}.
It should be pointed out that the comover dissociation channels
for charmonia are described in HSD with the proper individual thresholds
for each channel in contrast to the comover absorption model described in
Section II.A.

We recall that (as in Refs.
\cite{Cass01,Geiss99,Cass97,CassKo,brat03}) the charm degrees of
freedom are treated perturbatively and that initial hard processes
(such as $c\bar{c}$ or Drell-Yan production from $NN$ collisions) are
'precalculated' to achieve a scaling of the inclusive cross section
with the number of projectile and target nucleons as $A_P \times A_T$
when integrating over impact parameter $b$.

We typically perform 20 parallel runs for each impact parameter $b$ in
steps of $\Delta b$ = 0.5 fm from $b=0.5$ fm to $b = 2 R_T$, where
$R_T$ denotes the target radius.  Each parallel run here corresponds to
a single Au+Au collision event.  In central Au+Au collisions we have
$\sim$ 900 binary hard collisions at $\sqrt{s}$ = 17.3~GeV
(cf. Fig. 8 of \cite{Cass01}) and $\sim$ 1300 at $\sqrt{s}$ = 200~GeV.
In every binary collision we produce 1 particle for each species (i.e.
$J/\psi, \chi_c, \psi^\prime$, $D, \bar{D}, D^*, \bar{D}^*$, $D_s,
\bar{D}_s, D_s^*, \bar{D}_s^*$), however, with a different weight.
Thus, for 20 parallel runs we get about 1.8$\times 10^4$ (or 2.6$\times
10^4$) perturbative particles for each species.

For each single parallel run at fixed $b$ we obtain the final particle
multiplicity for all particle species as well as integral quantities
such as the transverse energy $E_T$ as a function of rapidity $y$, the
number of participants etc. Since we perform 20 parallel runs
simultaneously, the spread in the distributions of particle
multiplicities (or transverse energy) with respect to the individual
runs provides some information on the fluctuations of particle
multiplicities (as well as integral quantities). Vice versa, gating on
events with fixed transverse energy $E_T$ (in an interval $[E_T -\Delta
E_T/2, E_T +\Delta E_T/2]$) from all impact parameter $b$ we obtain a
distribution in the impact parameter $b$ that reflects the variation in
centrality for the selected event class. However, for the observables
presented in Section III we have checked that the fluctuations in
centrality have a minor impact on the normalized particle yields or
ratios.

The statistics is sufficiently good to reach an accuracy of particle
yields of a few percent in central collisions. This accuracy becomes
worth for peripheral collisions. Here we increase the number of
parallel runs in order to obtain approximately the same number of
charmonia and open charm mesons for fixed impact parameter as for
central collisions. Note, however, that the statistics also becomes
worth when including experimental acceptance cuts at SPS or RHIC
energies. Thus, when comparing to data, the overall accuracy is only on
the $\pm$ 5--7\% level. This is also due to the fact that only some
fraction of the initial charmonia survive the dynamical evolution due
to a large number of dissociation reactions (see below).

\section{Results for nucleus-nucleus collisions at SPS and RHIC}

\subsection{SPS energies}

\begin{figure}[t]
\centerline{\psfig{figure=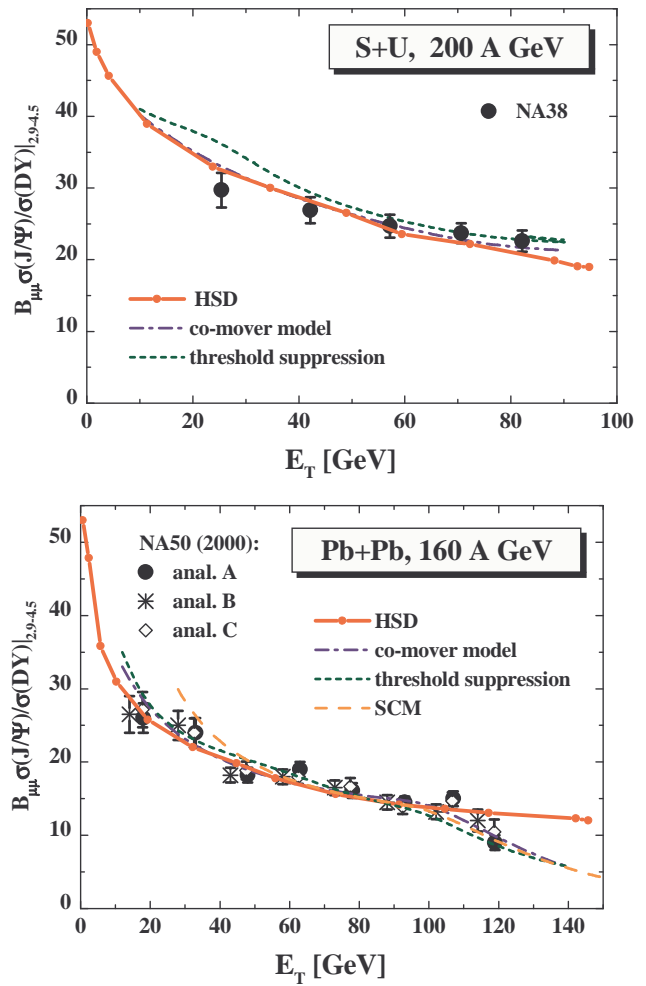,width=8.3cm}}
\caption{(Color online)
The ratio $B_{\mu\mu}\sigma(J/\psi) / \sigma(DY)$  as a function of the
transverse energy $E_T$ for S+U collisions at 200 A$\cdot$GeV (upper
part) and Pb+Pb collisions at 160 A$\cdot$GeV (lower part). The solid
lines with the full squares indicate the HSD results, the dashed-dotted
and short-dashed lines show the result of the suppression-only scenario
(comover and threshold suppression model, respectively) while the
long-dashed line stands for the statistical coalescence model (SCM).
The experimental data have been taken from Refs.
\protect\cite{NA38,NA50b,Ramello}.}
\label{Fig_JPSPS}
\end{figure}

Let us compare the charmonium suppression at SPS
energies  with experimental data from the NA50
Collaboration.  This collaboration presents its results
on $J/\psi$ suppression as the ratio
of the dimuon decay of $J/\psi$'s relative to the Drell-Yan background
in the 2.9 - 4.5 GeV invariant mass bin as a function of the transverse
energy $E_T$, i.e.
\begin{equation} \label{rat}
B_{\mu\mu}\sigma(J/\psi) / \sigma(DY)|_{2.9-4.5},
\end{equation}
where $B_{\mu\mu}$ is the branching ratio for $J/\psi\to \mu^+\mu^-$.

In the theoretical approaches we calculate the $J/\psi$ survival probability
$S_{J/\psi}$ defined as
\begin{equation}
\label{supp} S_{J/\psi} =
\frac{N^{J/\psi}_{fin}}{N^{J/\psi}_{BB}},
\end{equation}
where $N^{J/\psi}_{fin}$ and $N^{J/\psi}_{BB}$ denote the final number of
$J/\psi$ mesons and the number of $J/\psi$'s produced initially by
$BB$ reactions, respectively.
In order to compare our calculated results to experimental data we
need an extra input, i.e. the normalization factor
$B_{\mu\mu}\sigma_{NN}(J/\psi) / \sigma_{NN}(DY)$, which defines the
$J/\psi$ over Drell-Yan ratio for elementary nucleon-nucleon
collisions. We choose
$B_{\mu\mu}\sigma_{NN}(J/\psi) / \sigma_{NN}(DY) = 53$, in line
with a recent NA50 compilation \cite{NA50_03} obtained from
experiments on proton collisions with lighter targets
(cf. Sect. II.A.).

The experimental $\psi^\prime$ suppression is presented
by the ratio
\begin{equation} \label{pis}
\frac{B_{\mu\mu}(\psi^\prime\to \mu\mu)\sigma(\psi^\prime) }
{B_{\mu\mu}(J/\psi\to \mu\mu)\sigma(J/\psi) }.
\end{equation}
In our calculations we adopt this ratio to be 0.0165 for nucleon-nucleon
collisions, which is again based on the average over $pp, pd, pA$
reactions \cite{NA50_03}.

Fig. \ref{Fig_JPSPS}  shows the ratio $B_{\mu\mu}\sigma(J/\psi) /
\sigma(DY)$ as a function of the transverse energy $E_T$ for S+U
collisions at 200 A$\cdot$GeV (upper part) and Pb+Pb collisions at 160
A$\cdot$GeV (lower part). The solid line gives the HSD result
within the  comover absorption scenario for the cross sections defined
by (\ref{model}) while the various data points have been taken from
Refs.  \cite{NA38,NA50b,Ramello}. The dashed-dotted lines show results for
the comover absorption scenario  while the short
dashed lines stand for the threshold suppression model (cf. Sect. II.A.).
It is seen that all models are
compatible with the data for S+U as well as Pb+Pb, which is essentially
due the fit of the matrix elements $|M_i|^2$ in (6) for the transport
approach, $\sigma_{co}=1.0$~mb for the comover model and the threshold
participant densities $n_{\chi}=2.0$ fm$^{-2}$ and
 $n_{J/\psi}=3.8$ fm$^{-2}$ in the threshold suppression model.
The statistical coalescence model (long-dashed line in the lower
part of Fig. \ref{Fig_JPSPS}) also demonstrates a good agreement with
the data for (semi)central ($N_{part} > 100--150$) Pb+Pb collisions
(the S+U data are outside its domain of applicability) due to a fit
of the free parameters $\sigma^{NN}_{c\bar{c}}$ and $\xi_{\Delta y}$.

For the proper description of the drop of the ratio (\ref{rat}) in
Pb+Pb collisions at $E_T \approx$ 100 GeV one has to take into account
fluctuations of the transverse energy \cite{Capella2,Blaizot2}
and energy losses in the dimuon event sample \cite{Capella3,Kost_SPS2}.
To reproduce these effects in the transport approach, one would need
much better statistics, which is not feasible at present.

\begin{figure}[t]
\centerline{\psfig{figure=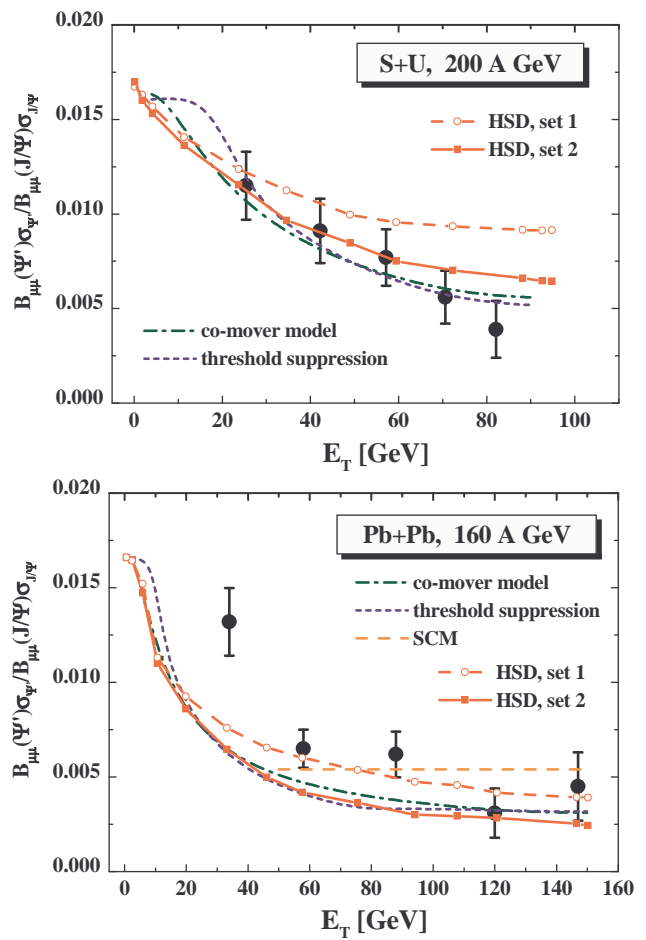,width=8.3cm}}
\caption{(Color online)
The ratio (\protect\ref{pis}) versus the transverse energy
$E_T$ for S+U collisions at 200 A$\cdot$GeV (upper part) and
Pb+Pb collisions at 160 A$\cdot$GeV (lower part).
The dashed lines with the open circles and the solid lines with the
full squares correspond to the HSD results calculated for two sets
of parameters for the $\psi^\prime$ matrix element - set 1 and set 2
(\protect\ref{set12}). The assignment of the other  lines is the same
as in Fig. \protect{\ref{Fig_JPSPS}}.
The experimental data have been taken from Refs. \protect\cite{NA38,Jaipur}.}
\label{Fig_PpSPS}
\end{figure}

The $\psi^{\prime}/J/\psi$ ratio  (\ref{pis}) is shown in
Fig. \ref{Fig_PpSPS}  versus the transverse energy
$E_T$ for S+U collisions at 200 A$\cdot$GeV (upper part) and
Pb+Pb collisions at 160 A$\cdot$GeV (lower part)
in comparison to the data from Refs. \cite{NA38,Jaipur}.
The dashed lines with the open circles and the solid lines with the
full squares correspond to the HSD results calculated for two sets
of parameters for the $\psi^\prime$ matrix element - set 1 and set 2
(\protect\ref{set12}). Here the results for 'set 1' overestimate
the ratio for S+U at high $E_T$, whereas they are compatible with
the ratio for Pb+Pb for $E_T \geq$ 60 GeV. The calculations for
'set 2' -- including a larger matrix element for $\psi^\prime$ --
systematically lead to a lower $\psi^{\prime}/J/\psi$ ratio as a
function of centrality. We note that no self energies for the $D,
\bar{D}$ mesons have been incorporated so far. The latter change
with baryon density and temperature and differ for $D$ and $\bar{D}$
mesons \cite{Amruta}. As pointed out in
Ref. \cite{Rappbrown} dropping $D, \bar{D}$ masses lead to an
increase of $J/\psi$ absorption by mesons and to a net lowering of
the $\psi^{\prime}/J/\psi$ ratio for central collisions.

The dashed-dotted lines in Fig. \ref{Fig_PpSPS} show the results for
the comover absorption scenario, the short dashed lines stand  for the
threshold suppression model, while the long dashed (constant) line
indicates the SCM results for Pb+Pb.  None of the models, however,
reproduces the ratio for $E_T \approx$ 35 GeV (for Pb+Pb).  All
approaches roughly yield a constant $\psi^{\prime}/J/\psi$ ratio for
Pb+Pb as a function of centrality for $E_T \geq$ 60 GeV.

\subsection{RHIC energies}

\begin{figure}[t]
\centerline{\psfig{figure=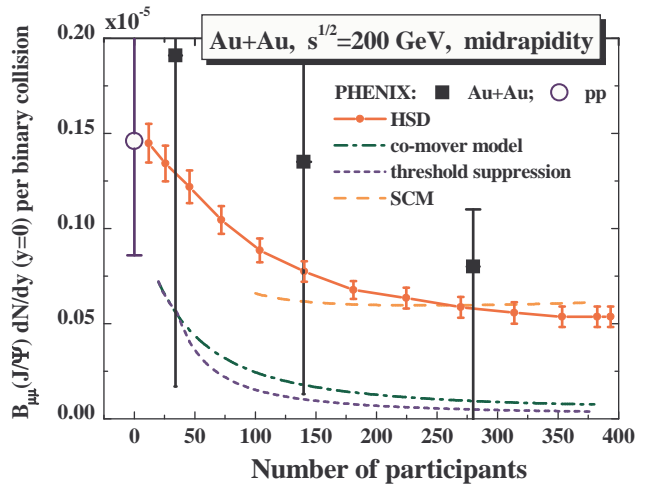,width=8.3cm}}
\caption{(Color online)
The calculated $J/\psi$ multiplicity per binary collision --
multiplied by the branching to dileptons --  as a function of the
number of participating nucleons $N_{part}$ in comparison to the
data from the PHENIX Collaboration \protect\cite{PHENIX_AA}
for $Au+Au$ and $pp$ reactions at $\sqrt{s}=200$ GeV.
The assignment of the lines is the same
as in Fig. \protect{\ref{Fig_JPSPS}}. }
\label{Fig_JPRHIC}
\end{figure}

Whereas the differences between the results of the models are rather
moderate at SPS energies due to a fit of the model parameters to the available data,
the situation changes substantially at
RHIC energies of $\sqrt{s}$ = 200 GeV. In Fig. \ref{Fig_JPRHIC} we show
the calculated $J/\psi$ multiplicity per binary collision --
multiplied by the branching to dileptons --  as a function of the
number of participating nucleons $N_{part}$ in comparison to the
data from the PHENIX Collaboration \cite{PHENIX_AA}
for $Au+Au$ and $pp$ reactions at $\sqrt{s}=200$ GeV.
The solid line with the full circles indicates the HSD results,
which roughly agree with  the SCM results (long dashed line) for
$N_{part} \geq$ 100.
The dashed-dotted line shows the results for the comover absorption scenario
while the short dashed line stands for the threshold suppression
model
(with the parameters fixed at SPS). It is seen that the comover
absorption model as well as the threshold  model lead to an almost
complete suppression of $J/\psi$ in central collisions (cf. also
Ref. \cite{Cass01}). As argued in Ref. \cite{brat03} this large
suppression in the comover model is essentially due to a neglect
of the backward channels $D+\bar{D} \rightarrow$ charmonia +
meson. In fact, the HSD calculations -- that include the various
backward channels -- lead only to a moderate $J/\psi$ suppression
roughly compatible with the result of the SCM. This finding might
suggest that the $J/\psi$ and open charm degrees of freedom
reach approximate chemical equilibrium for mid-central and central
Au+Au collisions at RHIC (see below). Unfortunately, the present data from
PHENIX do not allow to exclude any of the models so far.

\begin{figure}[t]
\centerline{\psfig{figure=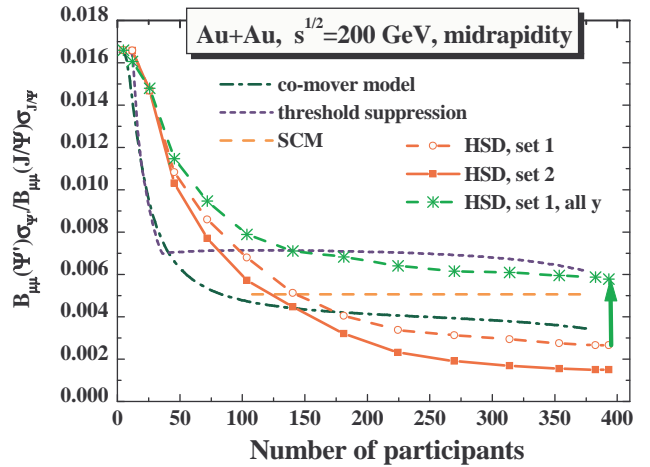,width=8.3cm}}
\caption{(Color online)
The midrapidity ratio (\protect\ref{pis})
 for Au+Au collisions at $\sqrt{s}=200$ GeV.
The dashed lines with the open circles and the solid lines with the
full squares correspond to the HSD results for two sets
of parameters for the $\psi^\prime$ matrix element - set 1 and set 2
(\protect\ref{set12}).
The assignment of the lines is the same as in the previous figures.
In addition, the
line with crosses displays the rapidity integrated ratio (\protect\ref{pis})
from HSD for set 1, which is even slightly above the SCM result for central
collisions.}
\label{Fig_PPRHIC}
\end{figure}

The $\psi^{\prime}/J/\psi$ ratio (\ref{pis}) at midrapidity
provides further information. It is displayed in Fig. \ref{Fig_PPRHIC}
for Au+Au collisions at $\sqrt{s}=200$ GeV versus the number of
participants $N_{part}$.
The dashed lines with the open circles and the solid lines with the
full squares correspond to the HSD results for two sets
of parameters for the $\psi^\prime$ matrix element - set 1 and set 2
(\protect\ref{set12}). These calculations give the lowest
$\psi^{\prime}/J/\psi$ ratio much below the ratio from the SCM
(long dashed line). The geometrical comover model (dashed-dotted line)
gives a higher ratio for central collisions than HSD. We attribute
this difference to the fact that in the geometrical comover model
only a single effective cross section
appears for all charmonia independent of threshold effects for
individual channels. The threshold suppression model (short dashed line)
provides the largest $\psi^{\prime}/J/\psi$ ratio for central
collisions even above the SCM results. Since the
predictions of the models differ by factors up to 4 future experiments
with high statistics should allow to exclude at least some of them.

In addition, we present in Fig. \ref{Fig_PPRHIC} (by the line with
crosses) the rapidity integrated ratio (\ref{pis})
from HSD for set 1, which is  slightly above the SCM result for central
collisions, however, still below the threshold model.
This finding clearly demonstrates that midrapidity and rapidity
integrated ratios have to be considered simultaneously before
conclusions on the amount of chemical equilibration can be
drawn.

\subsection{Quantitative analysis of reactions rates from HSD}

Whereas the results of the HSD transport approach for $J/\psi$
suppression show a rough agreement with the predictions from the
statistical coalescence model for $J/\psi$ (cf. Fig. \ref{Fig_JPRHIC}),
the $\psi^{\prime}$ to $J/\psi$ ratios differ considerably at
midrapidity (cf. Fig.
\ref{Fig_PPRHIC}). This demonstrates that a full chemical equilibrium
might not be achieved in the transport calculations since the total number of
$c, \bar{c}$ quarks are about the same in both models. In order to
understand these differences in more detail it is of interest to have a
closer look at the reaction rates from the HSD approach in total and
in a differential way with respect to rapidity.

\begin{figure}[t]
\centerline{\psfig{figure=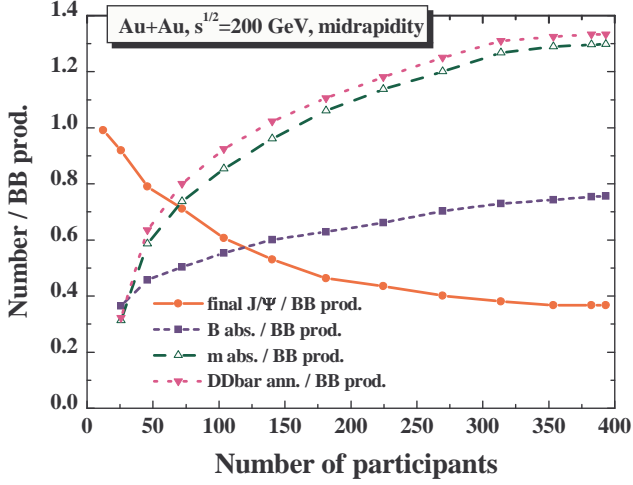,width=8.3cm}}
\caption{(Color online)
$J/\psi$ absorption and recreation by $D\bar D$ annihilation
versus $N_{part}$ for Au+Au collisions at midrapidity at $\sqrt{s}=200$ GeV.
The solid line with full circles shows the $J/\psi$ survival probability
(\protect\ref{supp}), i.e. the number of final $J/\psi$ mesons over the numbers
of $J/\psi$ initially produced by $BB$ reactions (denoted as '$BB$ prod.').
Note: in all cases here the $J/\psi$ numbers include the $J/\psi$
from the decays of $\chi_c$ and $\psi^\prime$. The dashed line with
full squares shows the integrated rate of $J/\psi$ absorption by baryons
(\protect\ref{rate2})
over '$BB$ prod.'. The dashed line with open triangles stands for the
rate of $J/\psi$ absorption by mesons (\protect\ref{rate3})
(normalized again to the primary '$BB$ prod.')
and the dotted line with full triangles shows the integrated rate of $J/\psi$
 recreated by $D\bar D$ annihilations (\protect\ref{rate4})
(normalized to '$BB$ prod.').}
\label{Fig_rateb}
\end{figure}

\begin{figure}[t]
\centerline{\psfig{figure=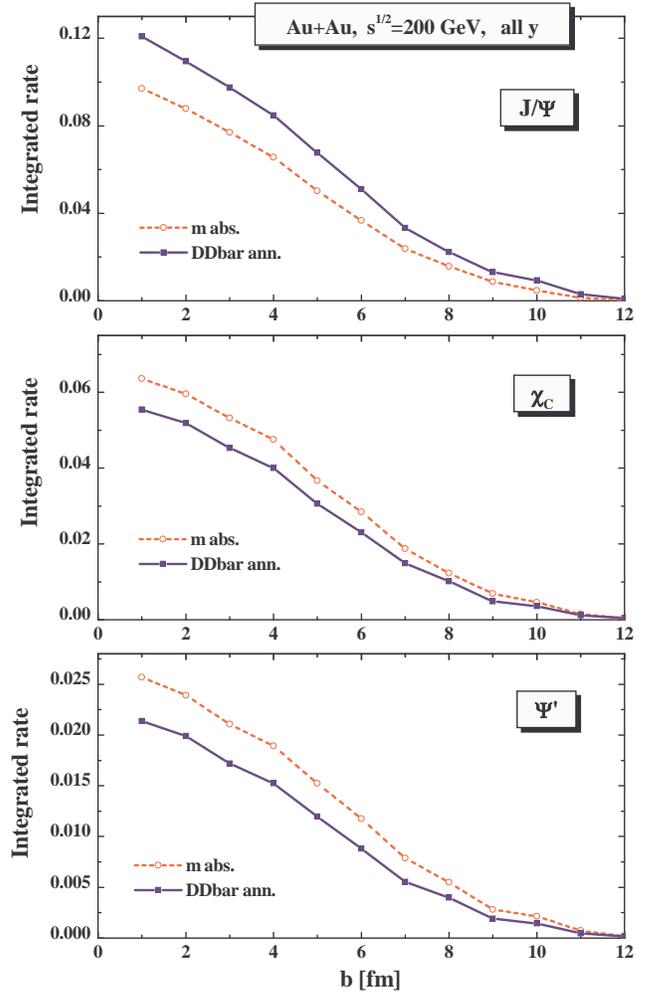,width=8.3cm}}
\caption{(Color online)
Integrated rate of $J/\psi$ (upper part), $\chi_c$ (middle part)
and $\psi^\prime$ (lower part) absorption by mesons (dashed lines
with open circles) in comparison to the recreation by $D\bar D$ annihilation
(solid line with full squares) as a function of the impact parameter $b$
for Au+Au at $\sqrt{s}$ = 200 GeV. }
\label{Fig_ratenp}
\end{figure}

To this aim we show in Fig. \ref{Fig_rateb} the $J/\psi$ absorption and
recreation by $D\bar D$ annihilation versus the number of participants
$N_{part}$ for Au+Au collisions at midrapidity for $\sqrt{s}=200$ GeV.
The solid line with full circles, furthermore,  shows the $J/\psi$
survival probability (\ref{supp}), i.e. the number of final $J/\psi$
mesons over the numbers of $J/\psi$ initially produced by $BB$
reactions (denoted as '$BB$ prod.').  We mention that the $J/\psi$
numbers here include the $J/\psi$ from the decays of $\chi_c$ and
$\psi^\prime$. The dashed line with full squares shows the integrated
rate of $J/\psi$ absorption by baryons
\begin{equation} \label{rate2}
\int_{-\infty}^\infty dt \frac{dN_{J/\psi + B \rightarrow X}}{dt}
\end{equation}
over the primary '$BB$ prod.'. As seen from Fig. \ref{Fig_rateb}
the final $J/\psi$ suppression is dominated by the dissociation with
baryons. The dashed line with open triangles indicates the integrated
rate of $J/\psi$ absorption by mesons
\begin{equation} \label{rate3}
\int_{-\infty}^\infty dt \frac{dN_{J/\psi + m \rightarrow D+\bar{D}}}{dt}
\end{equation}
(normalized again to the primary '$BB$ prod.') which is slightly
lower than the dotted line with full triangles, which stands for the
integrated rate of $J/\psi$
that are recreated by $D\bar D$ annihilations
\begin{equation} \label{rate4}
\int_{-\infty}^\infty dt \frac{dN_{D+\bar{D} \rightarrow J/\psi + m}}{dt}
\end{equation}
(normalized again to '$BB$ prod.'). Thus at practically all
centralities - except for very peripheral collisions -- the backward
reactions by $D+\bar{D}$ annihilation overcompensate the 'comover'
meson absorption. Nevertheless, both integrated rates are approximately
comparable suggesting an approximate  dynamical equilibrium between
charmonia, light mesons and open charm mesons (cf. Ref. \cite{brat03}).
However, a full chemical equlibrium is not achieved in the transport
calculations since the $\psi^\prime$ to $J/\psi$ ratio still depends on
the matrix element for the charmonium+meson coupling as seen explicitly
by comparing the results from set 1 with those from set 2.

The question remains why the $\psi^{\prime}/J/\psi$ ratios at
midrapidity differ significantly in comparison to the SCM. To shed some
light on this issue we show in Fig. \ref{Fig_ratenp} the time
integrated rate of $J/\psi$ (upper part), $\chi_c$ (middle part) and
$\psi^\prime$ (lower part) absorption by mesons (dashed lines with open
circles) in comparison to the recreation by $D\bar D$ annihilation
(solid line with full squares) as a function of the impact parameter
$b$. As already seen from Fig. \ref{Fig_rateb} the $J/\psi$ recreation
by $D+\bar{D}$ annihilation is larger than the $J/\psi$ dissociation
with mesons. This situation is inverse for the  $\chi_c$ (middle part)
and $\psi^\prime$ (lower part) and essentially related to the higher
mass of  $\chi_c$ and $\psi^\prime$ which lead to substantially larger
dissociation with pions due to the vicinity of the $D+\bar{D}$
threshold.  On the other hand the backward channels for  $\chi_c$ and
$\psi^\prime$ + meson recreation are suppressed by phase space relative
to the channel $J/\psi$ + meson. One note of caution has to be added
additionally: Due to sizeable differences in cross section the $\chi_c$
and $\psi^\prime$ dissociation extends to much larger times than the
backward recreation channels. Thus dynamically there is no common
freeze-out; the higher mass charmonium states ($\chi_c$ and
$\psi^\prime$) decouple at later times than $J/\psi$ and may only be
absorbed at late times but no longer recreated (see below).

\begin{figure}[t]
\centerline{\psfig{figure=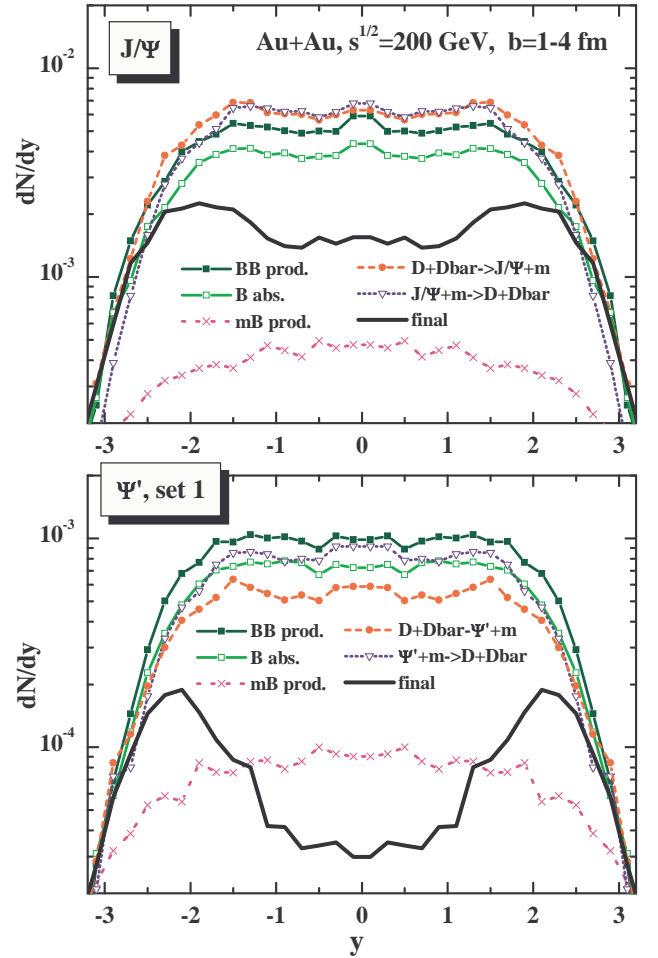,width=8.3cm}}
\caption{(Color online)
Rapidity distribution of $J/\psi$ (upper part)
and $\psi^\prime$ (calculated with set 1) production and absorption channels
for central (b= 1$\div$4 fm) collisions of Au+Au at $\sqrt{s}$ = 200 GeV                 .
The ordering of the different lines is as follows: the solid lines with
full squared stand for the rapidity distribution of $J/\psi$
($\psi^\prime$) mesons produced by initial $BB$ collisions while the
solid lines with open squared reflect the charmonia dissociation by
baryons ($B$ abs.); the dashed lines with crosses show the production
by $mB$ collisions. The dotted lines with open triangles show the
$J/\psi (\psi^\prime)$ dissociation by mesons while the dashed lines
with full circles stand for the recreation of charmonia by $D+\bar{D}$
annihilation. The full solid lines give the
final $J/\psi$ (upper part) and $\psi^\prime$ (lower part)
rapidity distributions. }
\label{Fig_yjprhic}
\end{figure}

Some further information on this issue is displayed in Fig.
\ref{Fig_yjprhic} where the rapidity distribution of the
individual channels is shown for $J/\psi$ (upper part) and
$\psi^\prime$ (lower part). Though all production and absorption
channels are rather flat in rapidity for -2 $\leq y \leq 2$ the
final $J/\psi$ rapidity distribution (thick solid line) shows a
local minimum for -1 $\leq y \leq 1$. This effect -- as a difference of
large numbers -- is related to the strong absorption with mesons and
recreation by open charm and anticharm mesons. Thus the absorption of
$J/\psi$ mesons relative to the initial production by baryon-baryon
collisions (full squares) shows a nontrivial rapidity dependence.
These effects are even more pronounced for the $\psi^\prime$ (lower
part) since the difference between the production and absorption
channels is most effectively seen around midrapidity -1 $\leq y \leq
1$, where the density of formed mesons is high and not very much
delayed by formation time effects.

\begin{figure}[t]
\centerline{\psfig{figure=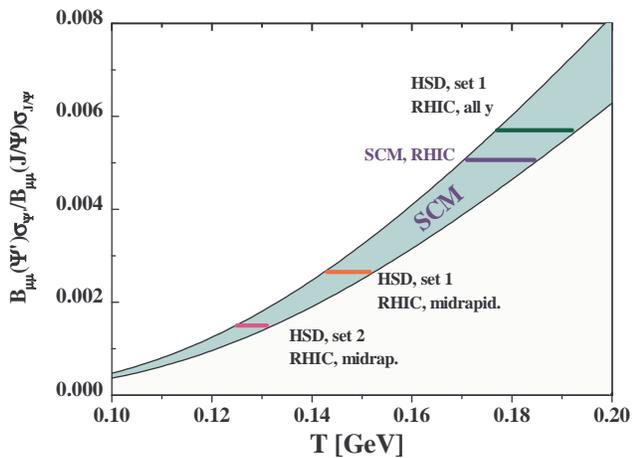,width=8.3cm}}
\caption{(Color online)
The ratio $\psi^{\prime}/J/\psi$ within the SCM as a function
of the temperature $T$. The upper and lower line of the shaded area
show the systematic uncertainty in the ratio that arises from
uncertainties in the branching ratios. The results for the midrapidity
ratios from HSD for set 1 and set 2 correspond to temperatures of
$\sim$ 150 and 130 MeV, whereas the SCM default result is quoted for
$T$= 177 MeV.  Note, that the rapidity integrated ratio from HSD for
set 1 corresponds to a temperature range from 175 to 190 MeV.}
\label{Fig_smsT}
\end{figure}

In addition we show in Fig. \ref{Fig_smsT} the $\psi^{\prime}/J/\psi$
ratio within the SCM as a function of the temperature $T$ at
freeze-out. The midrapidity ratios from HSD for set 1 and set 2
correspond to temperatures (in chemical equilibrium) of $\sim$ 150 and
130 MeV, respectively, whereas the SCM default result is displayed for
$T$=177 MeV. Thus in HSD the dynamical freeze-out conditions especially
for $\psi^\prime$ correspond to a later reaction phase than assumed in
the SCM. This result is plausible in view of the large reaction cross
sections for the channel $D+\bar{D} \leftrightarrow \psi^\prime$ +
meson.  Note, that the rapidity integrated ratio from HSD for set 1
corresponds to a temperature range from 175 to 190 MeV, which is
significantly higher than at midrapidity. Consequently, one has to
consider not only midrapidity ratios but their rapidity dependence as
well to obtain firm conclusions on freeze-out conditions.

\section{Summary}

In summarizing this work we have found that (in absence of open charm
enhancement in nucleus-nucleus collisions) the charmonium recreation by
the backward $D+\bar{D}$ channels plays no substantial role at SPS
energies, which leads to a good agreement between the comover and
threshold suppression models and the HSD transport calculations at this
energy.  However, the backward $D+\bar{D}$ channels become substantial
in Au + Au collisions at $\sqrt{s}$ = 200 GeV such that now an
approximate agreement of HSD with the statistical coalescence model is
achieved for the $J/\psi$ suppression in mid-central and central
collisions. We point out that a full chemical equilibration for the
hidden and open charm degrees of freedom is not achieved in the
transport calculations on the basis of hadronic interaction cross
sections since the $\psi^\prime$ to $J/\psi$ ratio still depends on the
matrix element for the $\psi^\prime$ coupling to mesons.  The latter
statement is solid since the cross sections employed for the $J/\psi,
\chi_c, \psi^\prime$ + meson$ \leftrightarrow D+ \bar{D}$ channels have
to be considered as upper limits because they are obtained from a fit
to the charmonium suppression data from NA50 at SPS energies by
discarding further absorption channels in a possibly pre-hadronic
phase.

In addition we have provided predictions for the $\psi^{\prime}/J/\psi$
ratio versus centrality, where the statistical coalescense model (SCM)
shows a larger value than the HSD approach at midrapidity. On the other
hand, rapidity integrated ratios in HSD are slightly higher than the
results from the SCM. This effect could be traced back to a significant
rapidity dependence of the final $\psi^\prime$ yield, since the net
$\psi^\prime$ absorption by mesons is maximal close to midrapidity.
These pronounced differences can be exploited in future measurements at
RHIC to distinguish a hadronic rescattering scenario from quark
coalescence close to the QGP phase boundary. \\

\section*{Acknowledgement}

The authors acknowledge inspiring discussions with
 P.~Braun-Munzinger and A. Mishra.
E.L.B. was supported by Deutsche Forschungsgemeinschaft (DFG) and GSI.
A.P.K. was supported by DFG.


\end{document}